\documentclass{article}

\usepackage{float}
\usepackage{arxiv}
\usepackage{subcaption}
\usepackage{multirow}
\usepackage{amsmath}
\usepackage[utf8]{inputenc} 
\usepackage[T1]{fontenc}    
\usepackage{hyperref}       
\usepackage{url}            
\usepackage{booktabs}       
\usepackage{amsfonts}       
\usepackage{nicefrac}       
\usepackage{microtype}      
\usepackage{lipsum}
\usepackage{graphicx}
\graphicspath{ {./images/} }
\title{Leveraging Time-Series Foundation Model for Subsurface Well Logs Prediction and Anomaly Detection}

\author{
    Ardiansyah Koeshidayatullah\textsuperscript{1,2*}, Abdulrahman Al-Fakih\textsuperscript{1*}, SanLinn Ismael Kaka\textsuperscript{1} \\
    \textsuperscript{1}Department of Geosciences, College of Petroleum Engineering and Geosciences, \\
    King Fahd University of Petroleum and Minerals, Dhahran, Saudi Arabia \\
    \textsuperscript{2}Center for Integrative Petroleum Research, College of Petroleum Engineering and Geosciences, \\
    King Fahd University of Petroleum and Minerals, Dhahran, Saudi Arabia \\
    *Corresponding author: \texttt{a.koeshidayatullah@kfupm.edu.sa}
}

\begin{document}
\maketitle
\begin{abstract}

The rise in energy demand has made it important to have suitable subsurface storage. , Achieving this requires detailed and accurate subsurface storage characterization, which often relies on the availability and quality of borehole well log data. However,  obtaining a complete suite of well-log data and ensuring high data quality is costly and time-consuming. In addition, missing data in well log measurement is common due to various reasons, such as borehole conditions and tool errors. While different machine learning and deep learning algorithms have been implemented to mitigate these issues with varying degrees of success, they often fall short in capturing the intricate, nonlinear relationships and long-term dependencies inherent in complex well log sequences. Furthermore, previous AI-driven models usually require retaining when new datasets are introduced and are limited to deployment in the same basin. Here, for the first time, we explored and evaluated the potential applications of the time-series foundation model, which leverages transformer architecture and a generative pre-trained approach to predict and detect anomalies in borehole well log data. Here, we fine-tuned and adopted TimeGPT architecture to analyze well log data. Our goal is to forecast different key log responses from various locations and detect anomalies in the logs with high accuracy. Our proposed model shows high-performance predictions, achieving correlation coefficient (\(R^2\)up to 87\% ) and a mean absolute percentage error (MAPE) as low as 1.95\%). In addition, the model’s capability, with a zero-shot approach, to detect anomalies in these logs has proven capable of identifying subtle yet critical deviations that could indicate drilling hazards or unexpected geological formations with an overall 93\% accuracy. Our model has demonstrated a significant advancement in both predictive accuracy and computational efficiency through its capability to perform zero-shot inference via a fine-tuning approach. Its application in well logs prediction not only enhances operational decision-making but also minimizes risks associated with subsurface exploration. 
\end{abstract}

\section*{Keywords}
Well Log; Time Series; Foundation Model; Geosciences; Reservoir

\section{INTRODUCTION}
{
Well logs are critical records in the oil and gas industry, capturing detailed measurements of subsurface properties along the depth of a wellbore. Not only does it provide a mathematical and physical measurement or representation of the rock properties in the subsurface, but also other surrounding information such as pressure or temperature can be recorded and deduced from well log data. These logs are essential for characterizing geological formations, estimating hydrocarbon reserves, and guiding drilling operations. Traditionally, well logs have been analyzed using statistical and traditional machine learning methods \cite{alfakih2023a, alfakih2023b, alfakih2023c}. However, recent advancements in time series forecasting have opened new avenues for well-log analyses \cite{pham2019, pham2020}. In particular, those incorporating deep learning techniques like recurrent neural networks (RNNs) and Long Short-Term Memory (LSTM) networks, are increasingly being used for this purpose due to their ability to capture temporal dependencies and non-linear relationships in the data \cite{wang2021, mohammed2024}. 

Time series forecasting models are particularly well-suited for well logs because these logs inherently represent sequential data over depth or time \cite{feng2021, mukherjee2024}. In addition to well log predictions, time-series-based analysis can be also utilized for detecting anomalies (e.g., abnormal pressure zones or equipment malfunctions) \cite{zhang2020, antariksa2023, altindal2024}. Early detection of such anomalies can prevent costly drilling issues and enhance operational safety. However, the application of these deep-learning-based models often requires extensive hyperparameter tuning and large amounts of training data, which can be computationally expensive and time-consuming. In addition, the models often fall short when dealing with complex, nonlinear patterns in the data, which is common in analyzing well-log responses.

Recently, the development of time-series foundation models, such as TimeGPT has revolutionized the forecasting process by utilizing Generative Pre-Trained transformers  \cite{garza2023}. Several studies have adopted this model for various tasks and demonstrated the superiority of this approach to other conventional methods \cite{liao2024, paroha2024}. Originally designed for natural language processing tasks, GPT models have shown remarkable versatility in sequence modeling, making them well-suited for time series data \cite{vaswani2017}. TimeGPT utilizes the self-attention mechanism inherent in Transformers to capture long-range dependencies in time series data, offering a scalable and efficient alternative to traditional and deep learning-based methods \cite{garza2023, liang2024, liu2024}. To date, despite the promising potential of such foundation models, no studies have explored their applications for well log predictions and anomaly detection. One major concern is the large volume of data (10\textsuperscript{6} to 10\textsuperscript{9}) to train the foundation models properly. Several works have indicated that fine-tuning foundation models with the dataset from the domain of interest can achieve satisfactory results even with limited data (10\textsuperscript{3})and harness the generalization capability of foundation models \cite{guo2024}.

Therefore, this study explores the effectiveness of the time-series foundation model, specifically by fine-tuning and adopting the TimeGPT model for well-log analysis of different hydrocarbon-prolific basins. Furthermore, we aim to establish a benchmark result for well-log prediction and anomaly detection by using a zero-shot approach. This is useful for future comparisons with other supervised, unsupervised, and foundation models that are applied for well-log analysis and interpretation.

\section{GENERATIVE AI FOR WELL LOG ANALYSIS}
The introduction of generative AI and transformer-based models for well-log data analysis and interpretation have opened new possibilities for not only generating synthetic data as pseudo-logs but also the efficiency, accuracy, and scope of well-log predictions. One of the first attempts at applying generative modeling to well-log data was through the autoregressive model. A study by \cite{kwon2020} showed that the deep learning autoregressive model surpasses human accuracy and with a more consistent interpretation in performing multiple petrophysical evaluations. Another work by \cite{kwon2020} leveraged conditional variational autoencoders for predicting sonic log responses and effectively considered uncertainty inherited from the measured data. The versatility of generative AI is further demonstrated by the work of \cite{qu2024}in performing data imputation for missing well-log data and achieving high accuracy and matches with the geological features. Furthermore, \cite{alfakih2024a} have successfully implemented sequence-based GAN to provide realistic synthetic well logs data and also fill the missing values with high accuracy. In addition, \cite{alfakih2024b} enhanced anomaly detection in well log data through the application of ensemble GANs.Recent works have also combined both transformer models and deep neural networks to enhance well-log interpretation and data processing, outperforming interpretation based on physical equations \cite{lin2023, xie2024}. While studies have highlighted the potential of generative AI, including time-series foundation models, in well-log data analysis, the applications remain limited. Hence, further exploration and implementation of such models are required to better understand the actual potential and limitations of generative modeling and foundation models for advancing well-log interpretation.

\section{METHODOLOGY}

\subsection{Dataset}
The datasets used in this study includes detailed well log from the North Sea Dutch region. The well logs include GR (Gamma Ray), DT (Sonic), NPHI (Neutron), and RHOB (Density) data, and cover a depth range of 1925-2065 meters, with a total of  6553 data points. These wells were selected due to their extensive logging data, which is pivotal for training and evaluating accurate machine-learning models.

\subsection{Computing Environment}
The analysis was conducted on a powerful workstation with 4x RTX A5500 24GB and RAM of 128 GB at the AI lab, King Fahd University of Petroleum and Minerals, Saudi Arabia. Python and the \texttt{NixtlaClient} API were used for forecasting and anomaly detection, along with a workflow specifically designed for handling time-series data for subsurface applications.

\subsection{Model Architecture}
The proposed model architecture is developed by adopting and fine-tuning the TimeGPT architecture \cite{garza2023}. It is built upon the GPT-2 architecture, which consists of multiple layers of Transformer blocks whereby each block includes multi-head self-attention mechanisms and position-wise feedforward networks \cite{radford2019}. The input to the model is a sequence of historical time series values, which are first embedded into a higher-dimensional space. Positional encodings are then added to the embedded sequence to incorporate the temporal order of the data \cite{vaswani2017}. This is particularly useful in time series forecasting, where understanding both short-term and long-term dependencies is crucial. Overall, the key components of our proposed model are the following (Figure~1):

\begin{enumerate}
    \item \textbf{Input Embedding Layer:} It consists of two components, time series data embedding and positional encoding. The embedding process is performed by transforming the time series data into a higher-dimensional space. This involves converting each data point in the sequence into a dense vector representation, capturing its temporal and contextual information. Since the transformer does not inherently understand the order of the input data, positional encodings are added to the input embeddings. This step introduces information about the position of each data point in the sequence, which is crucial for understanding the temporal relationships.

    \item \textbf{Multi-Head Self-Attention Mechanism:} The first procedure is to calculate the self-attention score. The self-attention mechanism allows the model to focus on different parts of the sequence when making predictions. Each element of the sequence is transformed into three vectors: Query (Q), Key (K), and Value (V). The model then calculates attention scores by taking the dot product of Q with K, normalizing them, and using these scores to weigh the V vectors. This process helps the model focus on the most relevant parts of the sequence. Finally, our model uses multiple attention heads to capture different aspects of the time series. Each head processes the data differently, allowing the model to consider various relationships within the sequence simultaneously. The outputs of all attention heads are then concatenated and linearly transformed to form the final output.

    \item \textbf{Feed-Forward Neural Network (FFN):} After the attention mechanism, the output is passed through a feed-forward neural network. This network consists of two linear layers with a ReLU activation in between. The FFN processes each position in the sequence independently, further transforming the data and introducing non-linearities.

    \item \textbf{Residual Connections and Layer Normalization:} To combat the vanishing gradient problem and help the model learn more efficiently, our proposed model includes residual connections that bypass the attention and feed-forward layers, adding the original input back to the output of these layers. In addition, after each attention and feed-forward block, layer normalization is applied. This helps stabilize the training process by normalizing the outputs, ensuring that the scale of the data remains consistent.

    \item \textbf{Output Layer:} The final output layer predicts the next value(s) in the time series. This layer can be adjusted to output a single value (univariate forecasting) or multiple values (multivariate forecasting), depending on the specific task.

    \item \textbf{Conformal Prediction:} Our model applies conformal prediction to provide uncertainty estimates through conformal prediction techniques. This allows the model to output prediction intervals, offering a confidence measure for the forecasts.
\end{enumerate}

The model is fine-tuned for time series forecasting by minimizing the mean squared error (MSE) between the predicted and actual values in the training set. Hyperparameters such as learning rate, batch size, and the number of attention heads were optimized using a grid search on the validation set. The model was trained using the Adam optimizer, with an initial learning rate of $1 \times 10^{-4}$ and a batch size of 64. Early stopping was employed to prevent overfitting, with training halted if the validation loss did not improve for 10 consecutive epochs. The evaluation was conducted on the test set (10\% of the total dataset collected from various basins).

\begin{figure}[htbp] 
    \centering
    \includegraphics[width=1.3\textwidth]{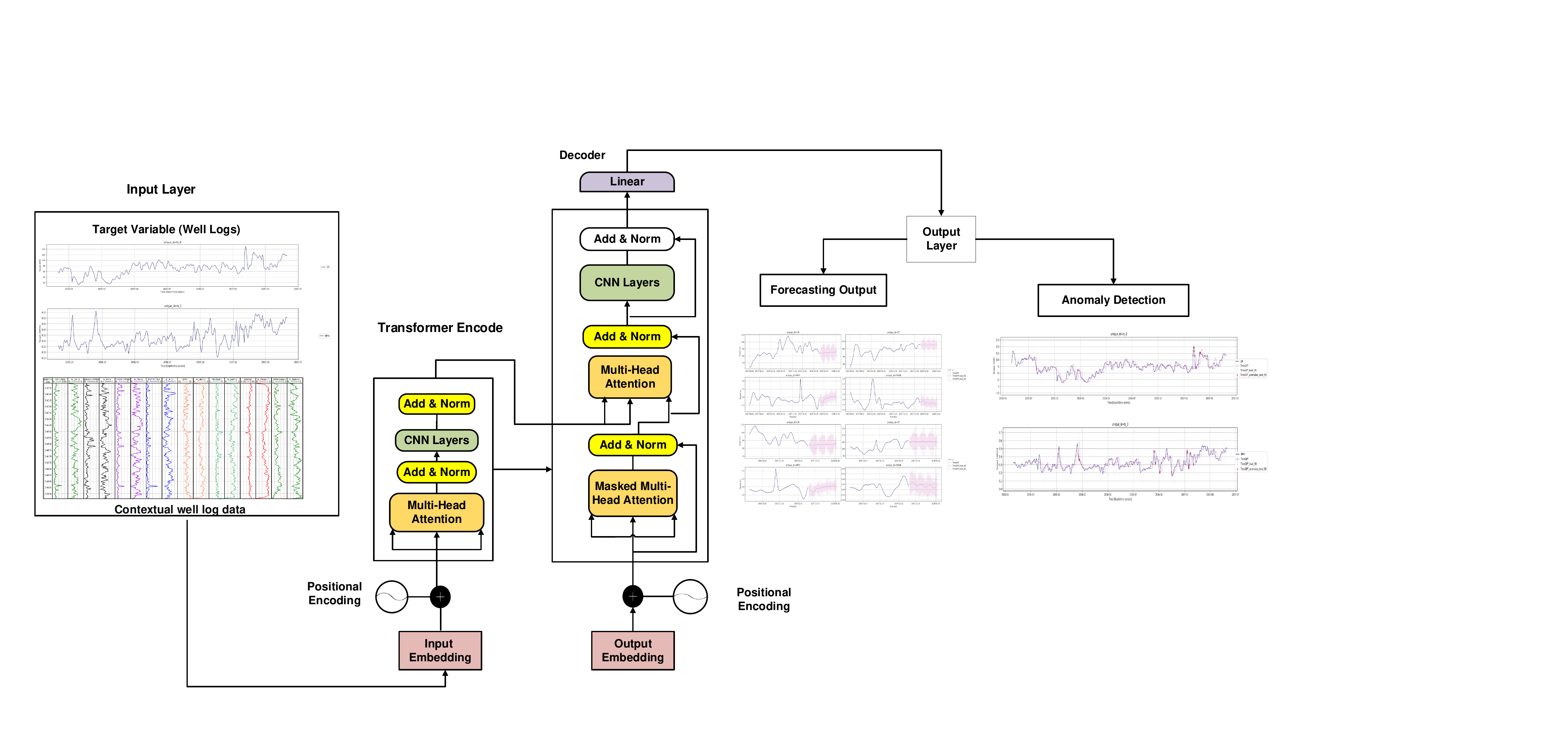}  
    \caption{The proposed workflow and architecture of fine-tuned time-series foundation model for well-log forecasting and anomaly detection.}
    
\end{figure}

\subsection{Evaluation Metrics}
\subsubsection{Well Log Prediction}
The performance of the models was assessed using several key metrics: $R^2$ (Coefficient of Determination), Mean Absolute Error (MAE), Root Mean Squared Error (RMSE), Mean Squared Error (MSE), and Mean Absolute Percentage Error (MAPE). These metrics provide complementary insights into the model’s accuracy and predictive capabilities.

\subsection*{Correlation Coefficient (Pearson's Correlation Coefficient)}
The correlation coefficient measures the strength and direction of the linear relationship between two variables:
\[
R^2 = \frac{\text{cov}(x, y)}{\sigma_x \cdot \sigma_y} \tag{1}
\]
where $\text{cov}(x, y)$ is the covariance between variables $X$ and $Y$, and $\sigma_x$ and $\sigma_y$ are the standard deviations of $X$ and $Y$, respectively. It ranges from 0 to 1, where a higher $R^2$ indicates a better fit, meaning the model explains a larger portion of the variance in the data.

\subsection*{Mean Absolute Error (MAE)}
The mean absolute error measures the difference between the predicted and actual values. It provides an easy-to-interpret measure of prediction accuracy, with lower MAE values indicating better performance:
\[
\text{MAE} = \frac{1}{n} \sum_{i=1}^n \left| y_i - \hat{y}_i \right| \tag{2}
\]
In which $y_i$ is the actual value, $\hat{y}_i$ is the predicted value, and $n$ is the number of data points.

\subsection*{Root Mean Squared Error (RMSE)}
The RMSE penalizes larger errors more than MAE and is often used to compare different models:
\[
\text{RMSE} = \sqrt{\frac{1}{n} \sum_{i=1}^n \left( y_i - \hat{y}_i \right)^2 } \tag{3}
\]

\subsection*{Mean Squared Error (MSE)}
The MSE provides the mean of the squared differences between actual and predicted values. Though it is less interpretable due to its squared units, it is widely used as a loss function in model optimization:
\[
\text{MSE} = \frac{1}{n} \sum_{i=1}^n \left( y_i - \hat{y}_i \right)^2 \tag{4}
\]
\subsubsection{Well Log Anomaly Detection}
To comprehensively evaluate the effectiveness of the models employed in this study, several key metrics were employed. These metrics offer insights into various facets of the model's predictive accuracy and reliability, particularly in the context of anomaly detection in well log data. The following metrics were used:

\subsection*{Accuracy (Acc)}
Accuracy quantifies the proportion of correctly predicted instances (both true positives and true negatives) out of all instances. It provides an overall indication of the model's correctness, as depicted in equation (5):
\[
\text{Accuracy} = \frac{\text{True Positives (TP)} + \text{True Negatives (TN)}}{\text{Total Instances}} \tag{5}
\]

\subsection*{Receiver Operating Characteristic - Area Under the Curve (ROC-AUC)}
This metric evaluates the model's capacity to differentiate between positive and negative classes across all potential thresholds. The ROC-AUC is represented by the area under the ROC curve, which plots the true positive rate against the false positive rate. A higher ROC-AUC signifies superior model performance.

\subsection*{Matthews Correlation Coefficient (MCC)}
The MCC offers a balanced measure that accounts for both true and false positives and negatives. It is particularly effective for assessing imbalanced datasets and provides a holistic assessment of binary classification quality, as depicted in equation (6):
\[
\text{MCC} = \frac{(\text{TP} \times \text{TN}) - (\text{FP} \times \text{FN})}{\sqrt{(\text{TP} + \text{FP})(\text{TP} + \text{FN})(\text{TN} + \text{FP})(\text{TN} + \text{FN})}} \tag{6}
\]

\subsection*{Confusion Matrix (CM)}
The CM offers a comprehensive breakdown of true positives, false positives, true negatives, and false negatives. It serves as a crucial tool for visualizing and assessing the performance of a classification algorithm. The CM provides insights into recall and precision, aiding in the evaluation of model performance.

\section{RESULTS AND INTERPRETATION}
\subsection{Exploratory Data Analysis}
Data analytics were conducted to understand the nature of the dataset and establish correlations between the input parameters. Based on the heatmap (Figure~2), key correlations between variables were revealed. A strong positive correlation (0.84) was observed between the RHOB and NPHI variables, followed by a moderate correlation (0.73) between the RHOB and GR. 
In contrast, the strongest negative correlation (-0.6) was observed between the NPHI and DT variables, as well as between the RHOB and DT (-0.55). The ILD variable exhibits weak correlations with others, indicating relative independence. These insights highlight significant relationships and potential multicollinearity between certain variables, which can inform feature selection and predictive modeling strategies in data analysis.

\begin{figure}[htbp] 
    \centering
    \includegraphics[width=1\textwidth]{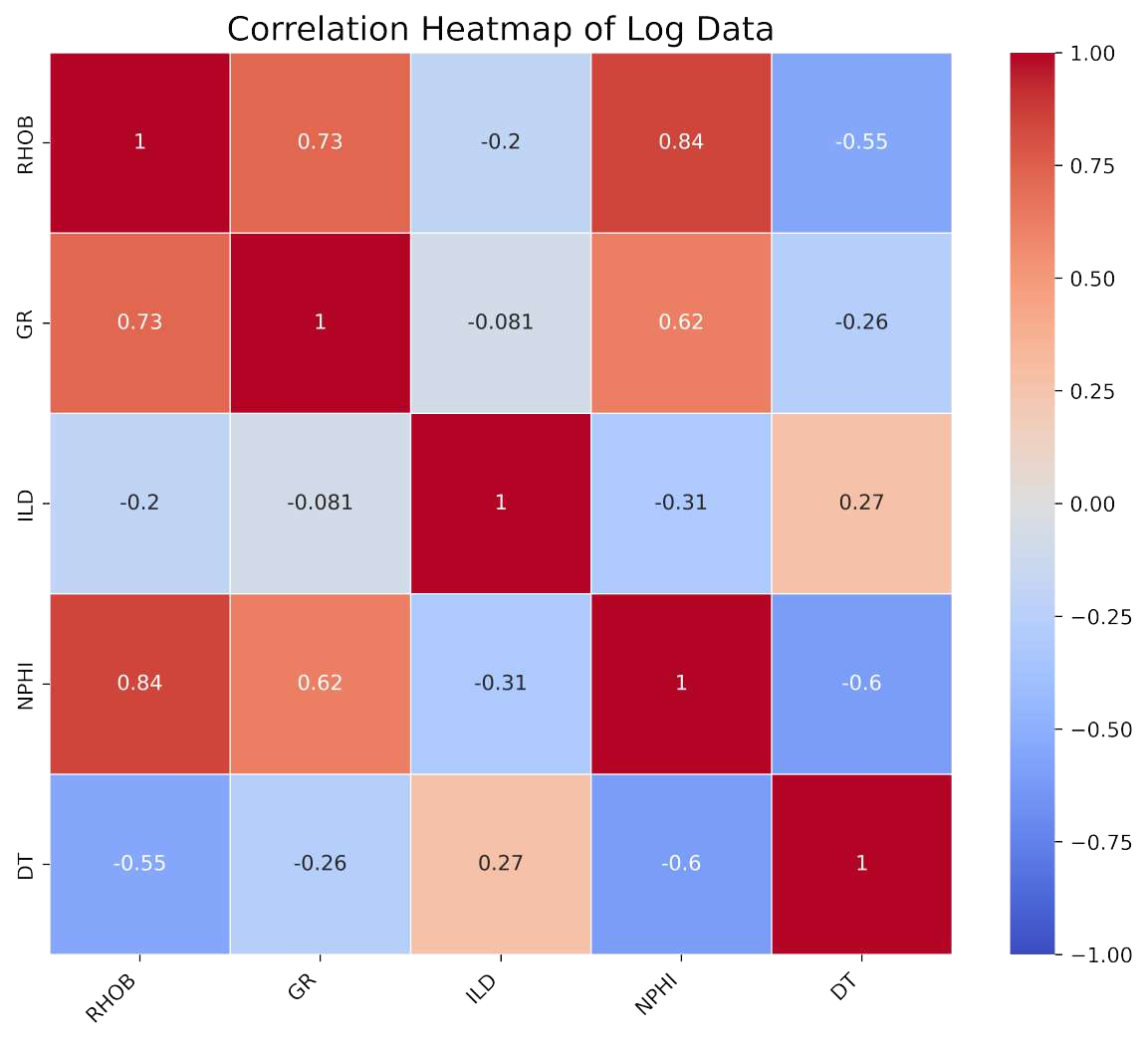}  
    \caption{Correlation heatmap of GR, DT, NPHI, ILD, and RHOB logs of the datasets.}
    
\end{figure}

The pair plot provides a detailed visualization of relationships among petrophysical properties (Figure~3): DT (sonic transit time), GR (gamma ray), ILD (deep resistivity), NPHI (neutron porosity), and RHOB (bulk density). Key trends include a negative correlation between DT and RHOB, suggesting lithological or porosity-related changes. GR shows distinct clusters, likely indicative of varying lithologies, and correlates negatively with ILD. NPHI exhibits a strong positive relationship with DT and negative trends with RHOB and ILD, reflecting porosity-lithology interactions. The distribution of ILD highlights high variability with sharp clusters, suggesting hydrocarbon zones or distinct rock types. These patterns provide insights into lithological and fluid property variations critical for reservoir characterization.

\begin{figure}[htbp] 
    \centering
    \includegraphics[width=1\textwidth]{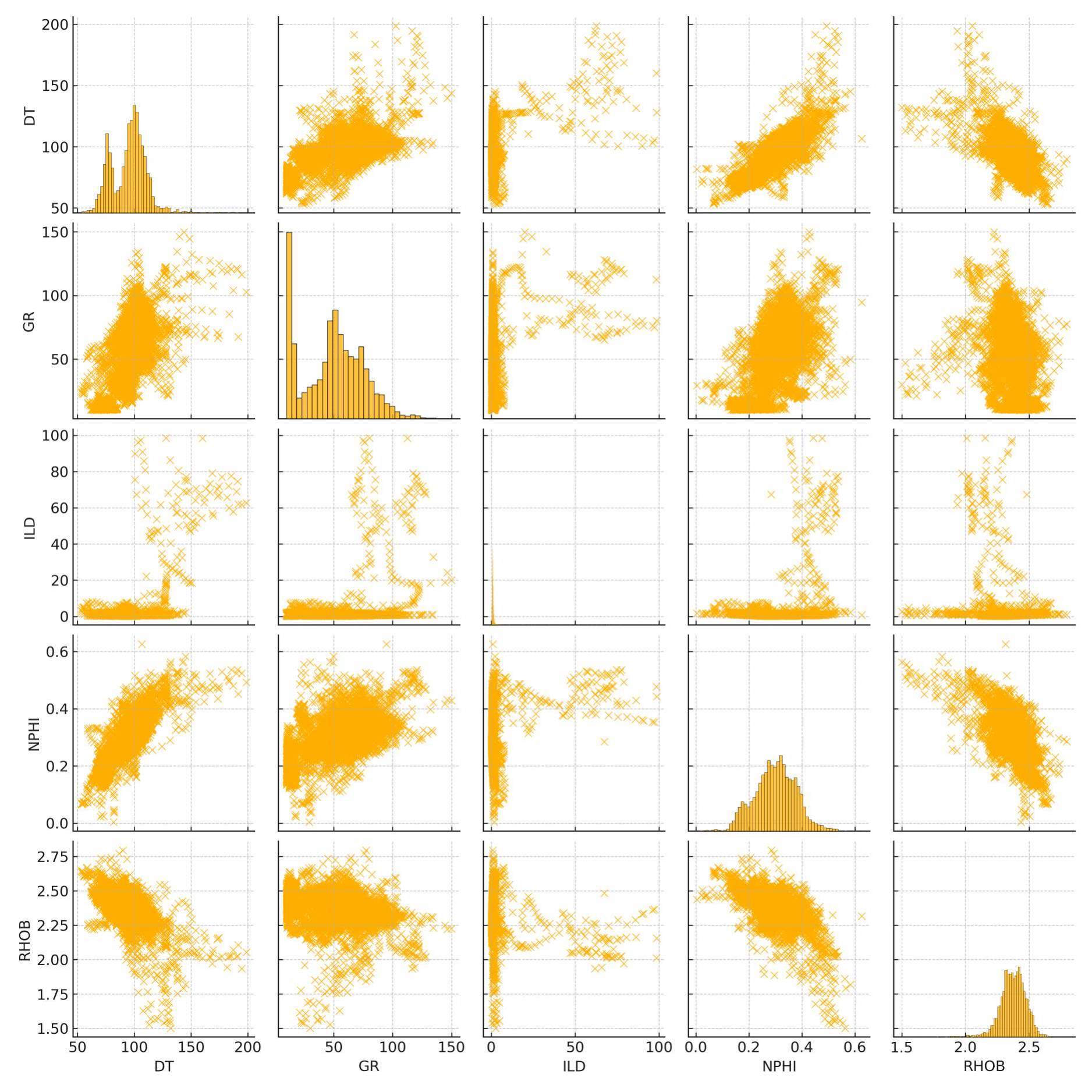}  
    \caption{Pair plot of GR, DT, NPHI, ILD, and RHOB logs from the training dataset.}
    
\end{figure}

\begin{table}[ht]
\centering
\renewcommand{\arraystretch}{1.2}
\caption{Evaluation metrics for well log forecasting}
\begin{tabular}{|l|c|c|c|c|c|}
\hline
\textbf{Log} & \textbf{MAE} & \textbf{RMSE} & \textbf{MAPE} & \textbf{R} & \textbf{$R^2$} \\ \hline
GR   & 3.201 & 4.343 & 5.572 & 0.932 & 0.865 \\ \hline
DT   & 2.659 & 3.736 & 2.539 & 0.823 & 0.645 \\ \hline
NPHI & 0.017 & 0.022 & 5.209 & 0.863 & 0.727 \\ \hline
RHOB & 0.033 & 0.046 & 1.395 & 0.741 & 0.477 \\ \hline
ILD  & 1.598 & 2.495 & 7.692 & 0.888 & 0.773 \\ \hline
\end{tabular}
\label{tab:evaluation_metrics}
\end{table}
\subsection{Well-Log Prediction}
In this study, various key log responses, including Gamma Ray (GR), Density (RHOB), Neutron (NPHI), Resistivity (ILD), and Sonic (DT), were evaluated to understand the potential applications of our proposed time-series foundation model. Among the different well-log data, the highest performance was obtained from forecasting GR, with the $R^2$ values achieving 87\% (Figure~4 and Table~1). In contrast, the model did not perform well in forecasting the RHOB logs, with $R^2$ values only around 48\% (Figure~4 and Table~1). The model exhibited comparable performance on the other log responses, with $R^2$ values ranging from 65\% to 77\% (Table~1). Furthermore, the model is capable of accurately forecasting log responses with drastic changes (high to low value) within a short distance(Figure~4). We also performed a future forecasting scenario, and the results showed that the model is able to reasonably predict the log responses beyond the measured depth.

\begin{figure}[htbp] 
    \centering
    \includegraphics[width=1\textwidth]{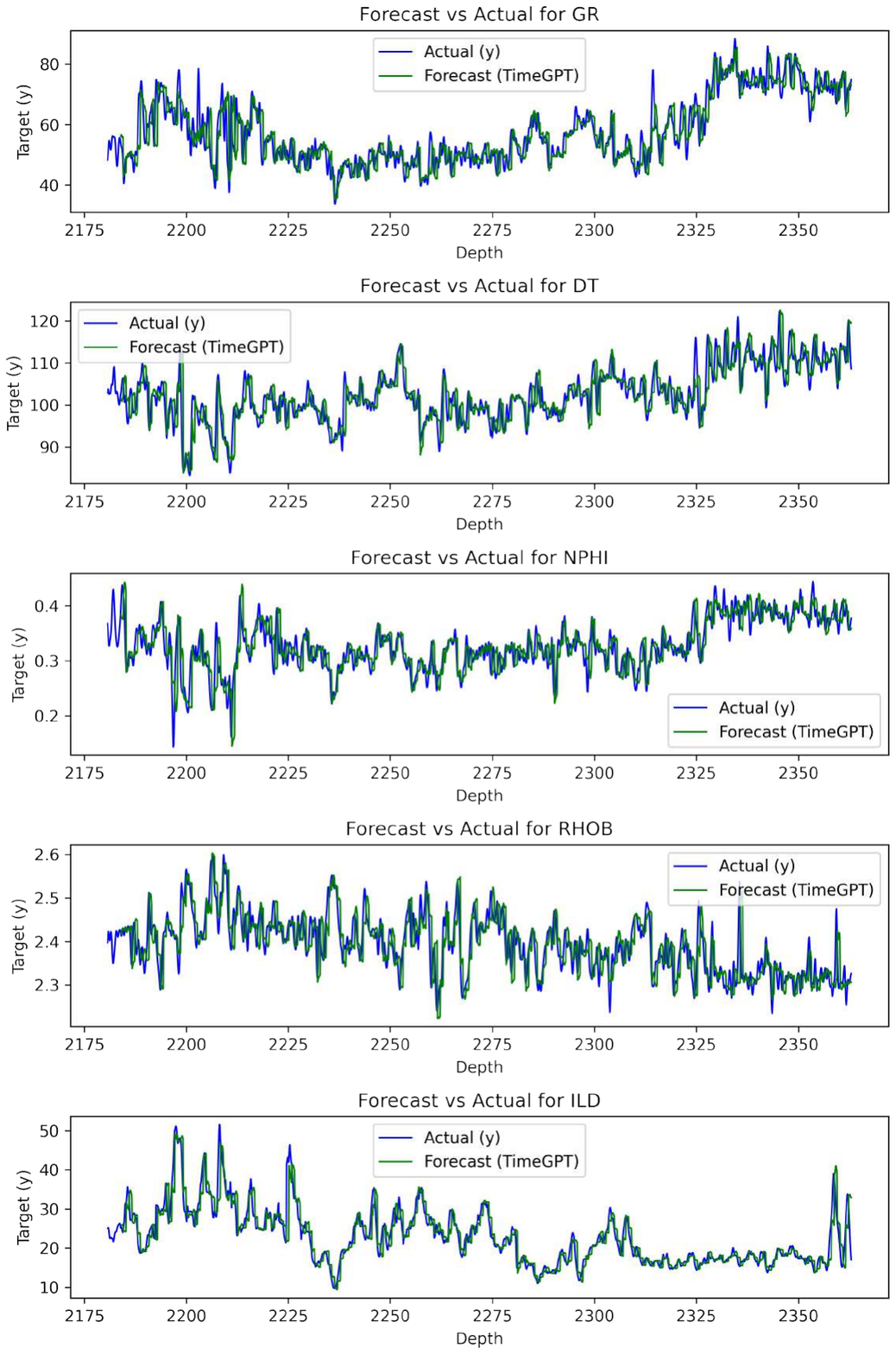}  
    \caption{Well log forecasting results across different log parameters show a good correlation between actual and forecast logs.}
    \end{figure}

\subsubsection*{Interpretation}
Forecasting log responses has been a focus of recent research in well prediction using machine learning \cite{mohammed2024}. ). In our study, the highest forecasting performance was achieved with the GR data. - The log responses reveal that the GR data is relatively dominantly charactherized by medium to high GR values, ranging from 60 to 80 API, with significant variability.  This indicates the highly heterogeneous nature of rock properties. Despite this variability, GR response is largely controlled by the radioactivity of the rock, making it more predictable. This predictability enables the model to identify patterns and achieve high prediction scores. In contrast, density log depends on multiple factors including lithology, mineralogy, porosity, and fluid composition. The complex interplay among these factors made density log prediction more challenging. Hence, additional data or measurement is needed to better capture the variations in density log responses.

\subsection{Well-Log Anomaly Detection}
Two different experiments were conducted for anomaly detection, using 90\% and 99\% confidence intervals. The model demonstrated consistent accuracy across the five different log responses, achieving 92\% for the 99\% confidence interval and 89\% for the 90\% interval (Figure~5 and Table~2). 
It is also noteworthy that the ROC-AUC values were similar across both log responses and confidence intervals, with values around 0.51 (Table~2).

\begin{figure}[htbp] 
    \centering
    \includegraphics[width=1.1\textwidth]{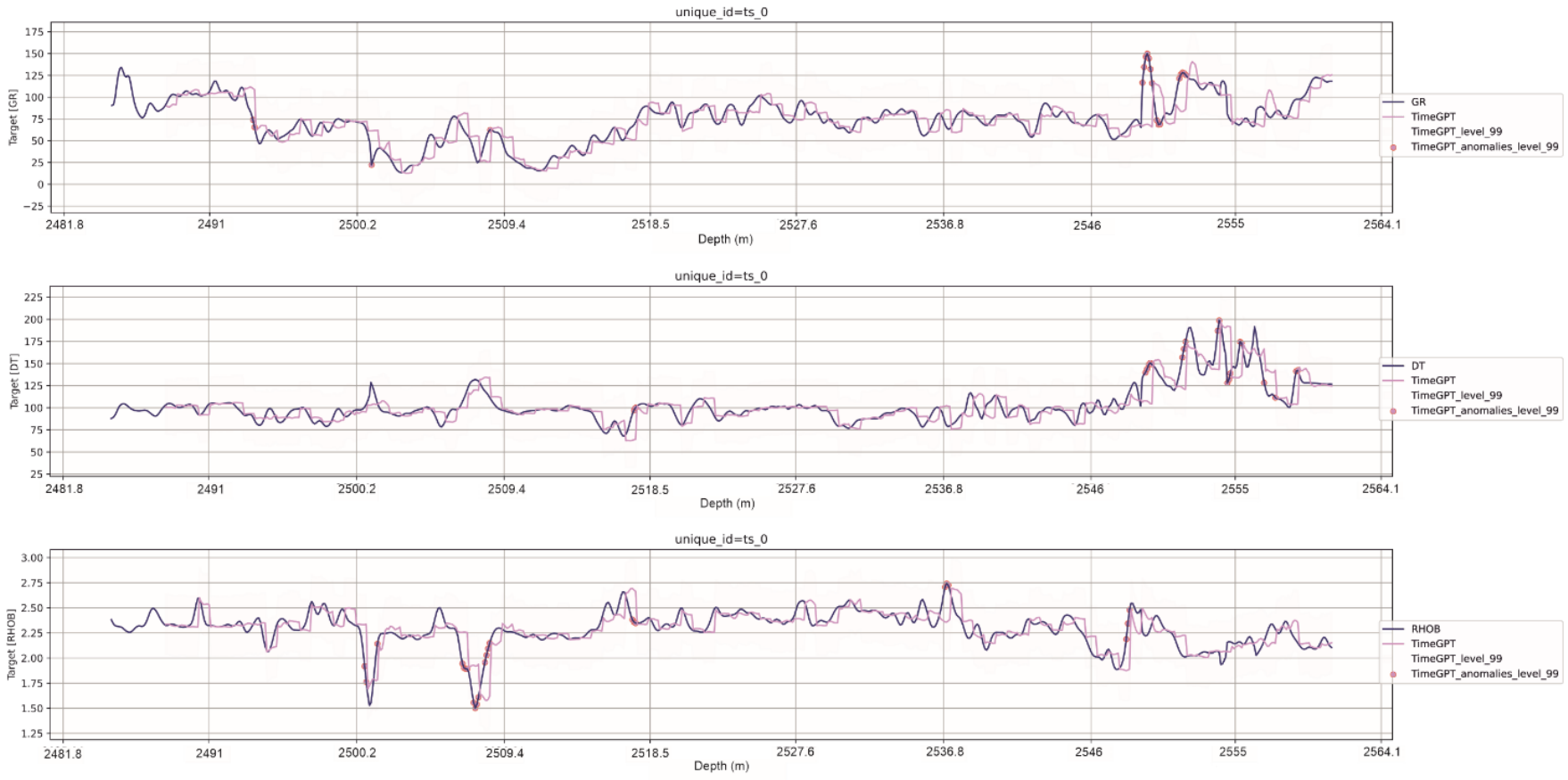}  
    \caption{Well log anomaly detection results across different log parameters show a good correlation between actual and forecast logs.}
    
\end{figure}

\subsubsection*{Interpretation}
Anomaly detection in well logs with machine learning is of great interest in optimizing log quality control and preventing misinterpretation \cite{struminskiy2019}. Our results from anomaly detection and the consistent metrics across different logs (Figure~5) suggest that the anomaly is most likely controlled by two factors: (i) related to geological heterogeneity (e.g., abrupt lithological changes or fluid composition). This can be observed where the anomalies occurred at different depths or only observed in one of the logs in the same depth, and (ii) the detected anomalies were sourced from either tool errors or bad borehole conditions as evident from the occurrence of anomaly across the log responses in the same depths (Figure~5). 
In addition, some of the log readings were exceptionally high or low which may indicate a tool error during the measurement. The consistency of the model to detect anomalies suggests that our approach is robust even with a zero-shot approach.

\begin{table}[htp]
\centering
\renewcommand{\arraystretch}{1.2}
\caption{Summary of performance metrics for various datasets for well-log anomaly detection}
\begin{tabular}{|l|c|c|c|c|}
\hline
\textbf{Dataset} & \textbf{Interval (\%)} & \textbf{Accuracy} & \textbf{ROC-AUC} & \textbf{MCC} \\ \hline
GR   & 99 & 0.9220 & 0.5031 & 0.0081 \\ 
     & 90 & 0.8826 & 0.5181 & 0.0268 \\ \hline
DT   & 99 & 0.9246 & 0.5155 & 0.0411 \\ 
     & 90 & 0.8926 & 0.5084 & 0.0133 \\ \hline
ILD  & 99 & 0.9222 & 0.5074 & 0.0194 \\ 
     & 90 & 0.9088 & 0.5042 & 0.0074 \\ \hline
RHOB & 99 & 0.9238 & 0.5096 & 0.0255 \\ 
     & 90 & 0.8945 & 0.5161 & 0.0257 \\ \hline
NPHI & 99 & 0.9209 & 0.5039 & 0.0100 \\ 
     & 90 & 0.8815 & 0.5126 & 0.0186 \\ \hline
\end{tabular}
\label{tab:anomaly_detection_metrics}
\end{table}

\section{DISCUSSIONS}
\subsection{Time-Series Foundation Models in Geosciences}
Generative AI models, including foundation models, are beginning to find applications in geosciences such as reservoir modeling, seismic data augmentation, and production optimization \cite{ferreira2022, koeshidayatullah2023}. These models assist in analyzing large volumes of geological and geophysical data, offering valuable insights to aid decision-making in reservoir engineering, drilling, and production management. However, the specific application of foundation models in these domains remains underexplored suggesting potential for future research.

In the broader context of time series forecasting, foundation models like TimeGPT have demonstrated superior performance. Their ability to handle multivariate forecasting tasks and support exogenous variables makes them adaptable to various data-driven scenarios. The model's incorporation of conformal prediction for estimating prediction intervals is particularly useful in anomaly detection, which could be highly beneficial in detecting irregular patterns in geoscience data. Focused research and adaptation of time-series foundation models in geosciences could unlock new capabilities, especially in predictive modeling and real-time data analysis. These advancements could be instrumental in applications such as monitoring environmental changes, predicting natural disasters, and optimizing resource extraction processes in petroleum engineering.

When compared with the traditional machine learning approach, the time-series foundation model offers several distinct advantages, notably its zero-shot approach, making it a powerful tool for a wide range of applications: (i) Our fine-tuned foundation model is designed to handle complex, multivariate time series data, where multiple variables might influence the target variable. Traditional methods like ARIMA or simple Exponential Smoothing models often struggle with multivariate data unless they are specifically designed for such tasks. In addition, as it is built on transformer architecture, the model naturally manages these complexities by modeling the interactions between different variables effectively. (ii) Unlike traditional models, which often have a limited window of past data they can consider (e.g., ARIMA and even many RNN-based methods), the proposed model can capture long-range dependencies. The transformer architecture enables the model to consider much longer sequences of past data, allowing it to recognize and leverage patterns that occur over extended periods. (iii) The foundation model can perform well with minimal or no specific training on a new dataset, thanks to its pre-training on vast amounts of time series data. This zero-shot or few-shot learning capability is a significant advantage when time or data is limited. Traditional models usually require substantial data and time to train on new tasks, whereas the model can be quickly adapted or even directly applied to new scenarios. Although many foundation models perform well out of the box, they can also be fine-tuned to specific datasets, allowing for further improvements in accuracy. This adaptability is less common in traditional models, which often require complete retraining when applied to new datasets or tasks. Lastly, the proposed model is highly scalable, making it suitable for large-scale datasets and real-time applications. The parallel processing capabilities of transformers allow TimeGPT to handle large datasets more efficiently than many traditional models, which can struggle with the computational load of large time series. The conformal prediction approach provides prediction intervals and thereby quantifies uncertainty in the forecasts. This is a significant advantage in critical applications where understanding the confidence of predictions is crucial, such as in financial forecasting or risk management. Traditional methods often provide point estimates without clear indications of their confidence levels, unless combined with additional statistical methods.

\subsection{Limitations and Future Recommendations}
While the time-series foundation model represents a significant advancement in the field of time-series forecasting, it is not without its limitations. These challenges need to be addressed to enhance the model's performance and applicability across different domains. This discussion will outline the current limitations of TimeGPT and provide recommendations for future research and development.

The Transformer architecture, which underpins TimeGPT architecture, is known for its computational intensity, particularly in terms of memory usage and training time. The self-attention mechanism in Transformers scales quadratically with the length of the input sequence, making it challenging to apply TimeGPT to long time series datasets without significant computational resources \cite{vaswani2017}. Due to its high capacity, foundation models are susceptible to overfitting, especially when trained on small or noisy datasets. Although techniques like dropout and early stopping are employed to mitigate this risk, overfitting remains a concern, potentially leading to poor generalization of unseen data \cite{hewamalage2023}. This limitation is particularly relevant in non-stationary time series where patterns can change over time, and the model may struggle to adapt to these shifts.

To address the computational complexity of GPT-based time-series foundation models, future research could focus on developing more efficient variants of the Transformer architecture. Techniques such as sparse attention mechanisms or low-rank approximations could reduce the computational burden without sacrificing accuracy \cite{shen2022,wang2024}. To mitigate the risk of overfitting and improve the model's adaptability to non-stationary time series, future iterations of time-series foundation models could incorporate adaptive learning mechanisms. For example, integrating online learning techniques that update the model as new data arrives could help the model maintain its performance in dynamically changing environments. Regularization strategies, such as ensemble learning or Bayesian methods, could also be explored to enhance the model's robustness against overfitting \cite{masini2023}.

\section{CONCLUSIONS}
The integration of time series forecasting and anomaly detection into well log analysis marks a significant advancement in subsurface exploration and drilling operations. These approaches enable more accurate predictions and provide early warnings of potential issues, enhancing decision-making and reducing operational risks. The case studies highlighted above demonstrate the effectiveness of these techniques in various aspects of well-log analysis, such as predicting lithology changes and detecting drilling anomalies. As these methods continue to evolve, they are expected to play an increasingly vital role in optimizing reservoir development and production in the oil and gas industry.   The versatility of our proposed fine-tuned model, combined with its ability to handle complex, multivariate, and long-range dependencies, makes it an invaluable resource for researchers and practitioners in numerous domains. Our study highlights the promising potential of a generative AI model for time series forecasting in geosciences and earth sciences across various domains, though its presence in these fields is still emerging.



\end{document}